\documentclass[pra,preprint,superscriptaddress,amssymb]{revtex4-1}
\usepackage{graphicx}
\usepackage{bm}
\usepackage{amsmath,amssymb,fancybox}
\usepackage{theorem}

\theorembodyfont{\normalfont}
\newtheorem{theorem}{Theorem}
\newtheorem{definition}{Definition}

\begin{document}
\begin{flushright}
YITP-20-97
\end{flushright}

\title{
Trusted center verification model and
classical channel remote state preparation
}
\author{Tomoyuki Morimae}
\email{tomoyuki.morimae@yukawa.kyoto-u.ac.jp}
\affiliation{Yukawa Institute for Theoretical Physics,
Kyoto University, Japan}
\affiliation{PRESTO, JST, Japan}
\author{Yuki Takeuchi}
\affiliation{NTT Communication Science Laboratories, NTT Corporation, Japan}

\begin{abstract}
The classical channel remote state preparation (ccRSP) is an
important two-party
primitive in quantum cryptography.
Alice (classical polynomial-time) and Bob (quantum
polynomial-time)
exchange polynomial rounds of classical messages,
and Bob finally gets random single-qubit states while Alice finally
gets classical descriptions of the states.
In [T. Morimae, arXiv:2003.10712], 
an information-theoretically-sound non-interactive
protocol for the verification of quantum computing
was proposed.
The verifier of the protocol is classical, but
the trusted center is assumed that sends random single-qubit states to
the prover and their classical descriptions to the verifier.
If the trusted center can be replaced with a ccRSP protocol
while keeping the information-theoretical soundness,
an information-theoretically-sound
classical verification of quantum computing is possible, 
which solves the long-standing open problem.
In this paper, we show that it is not the case unless
BQP is contained in MA.
We also consider a general verification protocol where
the verifier or the trusted center first sends quantum states to
the prover, and then the prover and the verifier exchange
a constant round of classical messages.
We show that the first quantum message transmission cannot
be replaced with a
ccRSP protocol while keeping the information-theoretical soundness
unless BQP is contained in AM.
Furthermore, we also study the verification with
the computational soundness.
We show that if a ccRSP protocol satisfies
a certain condition even against any quantum polynomial-time
malicious prover, the replacement of the
trusted center with the ccRSP protocol realizes 
a computationally-sound classical verification of quantum computing.
The condition is weaker than the verifiability of the ccRSP.
At this moment, however, there is no known ccRSP protocol
that satisfies the condition.
If a simple construction of such a ccRSP protocol is found, 
the combination of it
with the trusted center verification model provides another 
simpler and modular proof of the
Mahadev's result.
We finally show that the trusted center model and its variant
with the ccRSP have extractors for low-energy states.
\end{abstract}
\maketitle

\section{Introduction}
Whether quantum computing is classically verifiable or not is
one of the most important open problems in quantum information 
science~\cite{Gottesman,AharonovVazirani,Andru_review}.
If the soundness is the computational one, the Mahadev's 
breakthrough~\cite{Mahadev} solves the open problem affirmatively.
Or, if more than two provers, who are entangled but not allowed
to communicate with each other, 
are allowed, the information-theoretical soundness
is possible for a classical verifier~\cite{MattMBQC,Ji,RUV,Grilo,Coladangelo}.
In this paper, we focus on the single prover setup and
the information-theoretical soundness
(except for Secs.~\ref{sec:computational} and~\ref{sec:extractor}).
Furthermore, we require that the honest prover is
quantum polynomial-time, and therefore the well-known fact 
${\rm BQP}\subseteq{\rm IP}$ does not solve the open problem.

In Ref.~\cite{TC}, an information-theoretically-sound
non-interactive protocol for the verification of quantum
computing was proposed. In this protocol, the verifier
is classical, but 
the trusted center is assumed. The trusted center first sends
random BB84 states (i.e., $|0\rangle$, $|1\rangle$, 
$|+\rangle\equiv\frac{|0\rangle+|1\rangle}{\sqrt{2}}$,
and $|-\rangle\equiv\frac{|0\rangle-|1\rangle}{\sqrt{2}}$) 
to the prover, and their classical
descriptions to the verifier.
The prover
then
sends a classical message to the verifier.
The verifier finally does classical polynomial-time computing
to make the decision. (For details, see Ref.~\cite{TC}.
In Sec.~\ref{sec:TC} of this paper, we explain the protocol
for the convenience of readers.)

The classical channel
remote state preparation (ccRSP) is an important primitive in
quantum cryptography. It is a two-party protocol between
Alice and Bob where Alice is classical polynomial-time,
and Bob is quantum polynomial-time.
Alice and Bob exchange polynomial rounds of classical messages,
and Bob finally gets random single-qubit states
while Alice finally gets their classical descriptions.
The concept of the remote state preparation 
was first introduced in Ref.~\cite{Vedrancoherent}
in the context of blind quantum computing. 
Ref.~\cite{VedranRSP} studies
the remote state preparation in an abstract framework
for blind quantum computing. 
Computationally-secure ccRSP protocols have been constructed
under the standard assumption in cryptography
that the LWE is hard for quantum 
computing~\cite{BCMVV,AndruVidick,MetgerVidick,Qfactory}.

If the trusted center of the protocol of Ref.~\cite{TC}
can be replaced with a ccRSP protocol while keeping the
information-theoretical soundness,
the information-theoretically-sound
classical verification of quantum computing is possible,
which solves the open problem affirmatively.
In this paper, we show that it is not the case unless 
${\rm BQP}\subseteq{\rm  MA}$.
Because ${\rm BQP}\subseteq{\rm MA}$ is not believed to happen,
our result suggests that the trusted center cannot be replaced
with the ccRSP while keeping the information-theoretical soundness.
(Actually, what we obtain is a slightly stronger result,
${\rm BQP}\subseteq{\rm MA}_{\rm BQP}$,
where ${\rm MA}_{\rm BQP}$ is MA with honest quantum polynomial-time
Merlin. Because ${\rm MA}_{\rm BQP}\subseteq{\rm MA}$, we obtain
${\rm BQP}\subseteq{\rm MA}$.)

The no-go result can be shown even for
approximate ccRSP protocols where the prover and the verifier succeed
with some probability $p_{succ}$ even if the prover is honest,
and what the prover gets is close to the ideal state.

Replacing the trusted center of Ref.~\cite{TC} with the ccRSP is a natural
approach to
solve the open problem, but our result shows that it does not work.
It does not mean the impossibility of the (information-theoretically sound)
classical verification of
quantum computing, because there might be another approach,
but at this moment we do not know any promising approach.
(For example, the combination of the Fitzsimons-Kashefi (FK) protocol~\cite{FK} with the ccRSP will not
work, because the malicious unbounded prover can learn all trap information. 
See Appendix~\ref{app:learn}.)
On the other hand,
showing the impossibility of the 
(information-theoretically sound)
classical verification of quantum computing is also difficult, because
it means the separation between BQP and BPP.
(If we define ${\rm IP}_{\rm BQP}$ as the set of decision problems that are
verified by an IP protocol with an honest quantum polynomial-time
prover, we have ${\rm BPP}\subseteq{\rm IP}_{\rm BQP}
\subseteq{\rm BQP}$. Therefore,
${\rm IP}_{\rm BQP}\neq{\rm BQP}$ means
${\rm BPP}\neq{\rm BQP}$.)

We also consider a general verification protocol where
the verifier or the trusted center first sends quantum states to
the prover, and then the prover and the verifier exchange
a constant round of classical messages.
We show that the first quantum message transmission cannot
be replaced with a ccRSP protocol unless BQP is contained in AM.
(More precisely, what we actually obtain
is ${\rm BQP}\subseteq{\rm IP}_{\rm BQP}[const]$, where
$[const]$ means a constant round, but it leads to
${\rm BQP}\subseteq{\rm AM}$ because
${\rm IP}_{\rm BQP}[const]\subseteq{\rm IP}[const]
\subseteq{\rm AM}$.)

The second proof technique can also be applied to show
that replacing the trusted center in the protocol of Ref.~\cite{TC}
with the ccRSP is impossible unless ${\rm BQP}\subseteq{\rm AM}$,
but we can show a stronger result, namely, 
${\rm BQP}\subseteq{\rm MA}$,
by using the specific
structure of the protocol of Ref.~\cite{TC}.

We also study the verification with the
computational soundness.
We show that if a ccRSP protocol satisfies
a certain condition
even against any quantum polynomial-time
malicious prover, the replacement of the
trusted center of the protocol of Ref.~\cite{TC}
with the ccRSP protocol realizes 
a computationally-sound classical verification of quantum computing.
The condition is weaker than the verifiability of the ccRSP.
It was believed that the verifiability of a ccRSP is necessary
if it is used as a subroutine of a protocol of
the verification of quantum computing, but this result suggests
that it is not necessarily the case.
At this moment, however,
no ccRSP protocol is known that satisfies the condition.
If a ccRSP protocol that satisfies the condition
is constructed in a simple way,
the combination of it with the protocol of Ref.~\cite{TC}
provides another simpler and modular proof of the Mahadev's result.

The condition is satisfied in the protocol where the prover sends
quantum states to the verifier and the verifier does measurements.
It means that we can construct an off-line-quantum verification protocol
where the quantum message is sent from the prover to the verifier.

We also show that the trusted center model and its variant with
the ccRSP have extractors for low-energy states.
A quantum proof of quantum knowledge
was first introduced in Refs.~\cite{BroadbentGrilo,CVZ},
and a classical proof of quantum knowledge was introduced
in Ref.~\cite{VZ}.

Finally, let us mention a recent related work.
The paper~\cite{Atul} showed three results on the ccRSP
in the context of blind quantum computing.
First, they showed that
the ccRSP cannot be composable secure under the no-cloning theorem.
There is, however, a possibility that 
the BFK protocol~\cite{BFK} 
combined with
a ccRSP protocol 
is still composable secure.
Their second result is that it is not the case 
unless the no-signaling principle is violated.
Finally, they showed that the BFK protocol
combined with the Qfactory protocol~\cite{Qfactory}
satisfies the game-based security.

This paper is organized as follows.
In Sec.~\ref{sec:TC}, we review the verification protocol
of Ref.~\cite{TC}.
In Sec.~\ref{sec:result}, we show our first result,
and then in Sec.~\ref{sec:result2}, we show the second result
on the general setup.
We study the verification with the computational 
soundness in Sec.~\ref{sec:computational}.
We introduce the off-line-quantum verification protocol
with quantum communication from the prover to the verifier
in Sec.~\ref{sec:offline}.
We finally show the existence of extractors in Sec.~\ref{sec:extractor}.
The computational soundness is considered only in Sec.~\ref{sec:computational}
and Sec.~\ref{sec:extractor}.
In other sections, we implicitly assume
that the malicious prover is unbounded.

\section{The verification protocol of Ref.~\cite{TC}}
\label{sec:TC}
In this section, we review the verification protocol of Ref.~\cite{TC}.
The protocol is given in Fig.~\ref{protocol:TC}.
It was shown in Ref.~\cite{TC} that the protocol can verify
any BQP problem:
\begin{theorem}[Ref.~\cite{TC}]
\label{theorem:newposthoc}
For any promise problem $A=(A_{yes},A_{no})$ in BQP,
Protocol~\ref{protocol:TC} satisfies both of the following
with some $c$ and $s$ such that $c-s\ge\frac{1}{poly(|x|)}$:
\begin{itemize}
\item
If $x\in A_{yes}$, the honest quantum polynomial-time prover's
behavior makes 
the verifier accept with probability at least $c$.
\item
If $x\in A_{no}$,
the verifier's acceptance probability is at most $s$ for any
(even unbounded)
prover's deviation.
\end{itemize}
\end{theorem}

In Ref.~\cite{TC}, the completeness and the soundness are shown
by introducing virtual protocols where the prover teleports
quantum states to the verifier.
In Appendix~\ref{app:TC}, we give a direct proof of the
completeness and the soundness for the convenience of the readers.

\begin{figure}[h]
\rule[1ex]{\textwidth}{0.5pt}
\begin{itemize}
\item[0.]
The input is an instance
$x\in A$ of
a promise problem $A=(A_{yes},A_{no})$ in BQP,
and a corresponding $N$-qubit local Hamiltonian
\begin{eqnarray*}
{\mathcal H}\equiv\sum_{i<j}
\frac{p_{i,j}}{2}\Big(\frac{I^{\otimes N}+s_{i,j} X_i\otimes X_j}{2} 
+\frac{I^{\otimes N}+s_{i,j} Z_i\otimes Z_j}{2} 
\Big)
\end{eqnarray*}
with $N=poly(|x|)$
such that if $x\in A_{yes}$ then the ground energy is less than $\alpha$,
and if $x\in A_{no}$ then the ground energy is larger than $\beta$
with $\beta-\alpha\ge\frac{1}{poly(|x|)}$.
Here,
$I\equiv|0\rangle\langle0|+|1\rangle\langle1|$
is the two-dimensional identity operator,
$X_i$ is the Pauli $X$ operator acting on the $i$th qubit,
$Z_i$ is the Pauli $Z$ operator acting on the $i$th qubit,
$p_{i,j}>0$, $\sum_{i<j} p_{i,j}=1$, and $s_{i,j}\in\{+1,-1\}$.
\item[1.]
The trusted center uniformly randomly chooses
$(h,m_1,...,m_N)\in\{0,1\}^{N+1}$.
The trusted center sends $\bigotimes_{j=1}^N(H^h|m_j\rangle)$ 
to the prover.
The trusted center sends $(h,m)$ to the verifier,
where $m\equiv(m_1,...,m_N)\in\{0,1\}^N$.
\item[2.]
The prover does a POVM measurement 
$\{\Pi_{x,z}\}_{x,z}$
on the received state.
When the prover is honest,
the POVM corresponds to the teleportation
of a low-energy state $|E_0\rangle$ of the local Hamiltonian 
${\mathcal H}$ as if
the states sent from the trusted center are halves
of Bell pairs. 
The prover sends the measurement
result, $(x,z)$, to the verifier,
where $x\equiv(x_1,...,x_N)\in\{0,1\}^N$
and
$z\equiv(z_1,...,z_N)\in\{0,1\}^N$.
\item[3.]
The verifier samples $(i,j)$ with probability $p_{i,j}$,
and accepts if and only if $(-1)^{m_i'}(-1)^{m_j'}=-s_{i,j}$,
where
$m_i'\equiv m_i\oplus (hz_i+(1-h)x_i)$.
\end{itemize} 
\rule[1ex]{\textwidth}{0.5pt}
\caption{The verification protocol of Ref.~\cite{TC}.}
\label{protocol:TC}
\end{figure}

\section{Replacement of the trusted center}
\label{sec:result}
Let us consider Protocol~\ref{protocol:ccRSP},
which is the same as Protocol~\ref{protocol:TC}
except that the trusted center is replaced with a ccRSP protocol.
As a ccRSP, we consider an approximate one:
if the prover behaves honestly, the verifier and the prover
succeed with probability
$p_{succ}$. If they are successful, the verifier gets
$(h,m)\in\{0,1\}^{N+1}$ and the prover gets 
an $N$-qubit state $\sigma_{h,m}$ with probability
$P(h,m)$, where
\begin{eqnarray*}
\frac{1}{2}\Big\| \sum_{h,m}P(h,m)\sigma_{h,m}-
\frac{1}{2^{N+1}}\sum_{h,m}\bigotimes_{j=1}^N H^h|m_j\rangle\langle m_j|H^h
\Big\|_1\le \epsilon
\end{eqnarray*}
is satisfied for a certain small $\epsilon$.
Even if the prover behaves honestly, 
they fail with probability $1-p_{succ}$. 
Furthermore, we assume that $p_{succ}$ is samplable in classical
polynomial-time, which is a reasonable assumption because
the description of the ccRSP protocol is known to the verifier.

We show that such a modified protocol is not
an information-theoretically-sound verification protocol
unless ${\rm BQP}\subseteq{\rm MA}_{\rm BQP}$.

Before stating the result, let us define the class
${\rm MA}_{\rm BQP}$.
\begin{definition}
A promise problem $A=(A_{yes},A_{no})$ is in ${\rm MA}_{\rm BQP}$ if
and only if there exists a classical probabilistic polynomial-time
verifier such that
\begin{itemize}
\item
If $x\in A_{yes}$, there exists a quantum polynomial-time prover
that sends a classical polynomial-length bit string to the verifier
such that the verifier accepts with probability at least $\frac{2}{3}$.
\item
If $x\in A_{no}$, for any polynomial-length classical bit string from
the prover (who can be unbounded), the verifier's
acceptance probability is at most $\frac{1}{3}$.
\end{itemize}
\end{definition}

It is easy to show that ${\rm MA}_{\rm BQP}\subseteq{\rm MA}$.
Now let us show our first result.

\begin{theorem}
Assume that Protocol~\ref{protocol:ccRSP}
can verify any BQP problem.
It means that
for any promise problem $A=(A_{yes},A_{no})$ in BQP,
Protocol~\ref{protocol:ccRSP} satisfies both of the following
with some $c$ and $s$ such that $c-s\ge\frac{1}{poly(|x|)}$:
\begin{itemize}
\item
If $x\in A_{yes}$, the honest quantum polynomial-time prover's
behavior makes the verifier
accept with probability at least $c$.
\item
If $x\in A_{no}$,
the verifier's acceptance probability is at most $s$ for any
(even unbounded)
prover's deviation.
\end{itemize}
Then, ${\rm BQP}\subseteq{\rm MA}_{\rm BQP}$.
\end{theorem}

\begin{figure}[h]
\rule[1ex]{\textwidth}{0.5pt}
\begin{itemize}
\item[0.]
The same as the step 0 of Protocol~\ref{protocol:TC}.
\item[1.]
The verifier and the prover run a ccRSP protocol.
If the prover behaves honestly,
they succeed with probability $p_{succ}$.
If they are successful,
the verifier gets
$(h,m_1,...,m_N)\in\{0,1\}^{N+1}$
and
the prover gets an $N$-qubit state $\sigma_{h,m}$ with probability
$P(h,m)$.
If they fail, the verifier rejects.
\item[2.]
The same as the step 2 of Protocol~\ref{protocol:TC}.
\item[3.]
The same as the step 3 of Protocol~\ref{protocol:TC}.
\end{itemize} 
\rule[1ex]{\textwidth}{0.5pt}
\caption{The modified protocol.}
\label{protocol:ccRSP}
\end{figure}

Before showing a proof, there is a remark. 
It is clear from the proof that what we require for the
ccRSP is only the (approximate) correctness. Neither the blindness nor 
the verifiability is required:
The correctness means that Alice and Bob get correct
outputs when they are honest. In the present case,
the correct outputs
are $\bigotimes_{j=1}^NH^h|m_j\rangle$ for Bob
and uniformly random $(h,m)\in\{0,1\}^{N+1}$ for Alice.
Usually when we use a ccRSP, we require the blindness or the
verifiability. The blindness means that $h$ or $m$ are
``hidden" to even malicious Bob, and the verifiability means that
even if Bob is malicious Alice can guarantee that Bob gets
the correct state (up to Bob's operation).
Our theorem requires the ccRSP to satisfy
only the minimum requirement, namely,
the correctness. (Furthermore, not the exact correctness,
but the approximate correctness is enough.)

{\it Proof}.
Let $A=(A_{yes},A_{no})$ be any BQP promise problem.
For any yes instance $x\in A_{yes}$, the verifier's
acceptance probability $p_{acc}^{honest}(x)$
of Protocol~\ref{protocol:ccRSP}
is 
\begin{eqnarray}
p_{acc}^{honest}(x)&=&
p_{succ}\sum_{h,m}P(h,m)\sum_{x,z}\mbox{Tr}(\Pi_{x,z}\sigma_{h,m})
\sum_{i<j}p_{i,j}\frac{1-s_{i,j}(-1)^{m_i'+m_j'}}{2}\nonumber\\
&=&
p_{succ}
\sum_{x,z}\mbox{Tr}[\Pi_{x,z}
\sum_{h,m}P(h,m)
\sigma_{h,m}]
\sum_{i<j}p_{i,j}\frac{1-s_{i,j}(-1)^{m_i'+m_j'}}{2}\nonumber\\
&\le&
p_{succ}
\sum_{x,z}\mbox{Tr}\Big[\Pi_{x,z}
\frac{1}{2^{N+1}}\sum_{h,m}
(H^{\otimes N})^h|m\rangle\langle m|(H^{\otimes N})^h\Big]
\sum_{i<j}p_{i,j}\frac{1-s_{i,j}(-1)^{m_i'+m_j'}}{2}+\epsilon\nonumber\\
&=&
p_{succ}[
1-\mbox{Tr}({\mathcal H}|E_0\rangle\langle E_0|)]+\epsilon,
\label{yes}
\end{eqnarray}
where $|m\rangle\equiv\bigotimes_{j=1}^N|m_j\rangle$.
For the last equality, see Appendix~\ref{app:TC}.

Let $x\in A_{no}$ be any no instance.
Let us consider the following malicious unbounded
prover's attack:
\begin{itemize}
\item[1.]
When the prover and the verifier run the ccRSP protocol,
the prover classically simulates prover's honest quantum behavior.
(The verifier cannot distinguish whether the prover is really
doing the honest quantum procedure or
simulating it classically. See Appendix~\ref{app:learn}.)
If they are successful,
the verifier gets $(h,m)\in\{0,1\}^{N+1}$ with probability $P(h,m)$.
The prover can learn $(h,m)$ 
because the prover has the classical description of
$\sigma_{h,m}$.
(See Appendix~\ref{app:learn}.)
\item[2.]
If $h=0$, the prover chooses $(x,z)\in\{0,1\}^N\times\{0,1\}^N$,
where $x$ is sampled from a certain distribution $D$,
and $z$ is uniformly randomly chosen.
The prover sends $(x\oplus m,z)$ to the verifier.
Here, $x\oplus m\equiv(x_1\oplus m_1,...,x_N\oplus m_N)$.
If $h=1$, the prover chooses $(x,z)\in\{0,1\}^N\times\{0,1\}^N$,
where $z$ is sampled from the distribution $D$,
and $x$ is uniformly randomly chosen.
The prover sends $(x,z\oplus m)$ to the verifier.
Here, $z\oplus m\equiv(z_1\oplus m_1,...,z_N\oplus m_N)$.
\end{itemize}
The verifier's acceptance probability 
$p_{acc}^{malicious}(x)$ under this prover's attack is
\begin{eqnarray}
p_{acc}^{malicious}(x)
&=&
p_{succ}\sum_mP(0,m)\sum_{x,z}\frac{1}{2^N}D(x)
\sum_{i<j}p_{i,j}\frac{1-(-1)^{m_i+(x_i+m_i)+m_j+(x_j+m_j)}s_{i,j}}{2}\nonumber\\
&&+p_{succ}\sum_mP(1,m)\sum_{x,z}\frac{1}{2^N}D(z)
\sum_{i<j}p_{i,j}\frac{1-(-1)^{m_i+(z_i+m_i)+m_j+(z_j+m_j)}s_{i,j}}{2}\nonumber\\
&=&p_{succ}\sum_mP(0,m)\sum_{x,z}\frac{1}{2^N}D(x)
\sum_{i<j}p_{i,j}
\langle x|\frac{I^{\otimes N}-s_{i,j}Z_i\otimes Z_j}{2}
|x\rangle\nonumber\\ 
&&+p_{succ}\sum_mP(1,m)\sum_{x,z}\frac{1}{2^N}D(z)
\sum_{i<j}p_{i,j}
\langle z|\frac{I^{\otimes N}-s_{i,j}Z_i\otimes Z_j}{2}
|z\rangle\nonumber\\
&=&p_{succ}
\mbox{Tr}\Big[(I^{\otimes N}-{\mathcal H}_Z)
\sum_{k\in\{0,1\}^N}D(k)
|k\rangle\langle k|\Big]\label{no},
\end{eqnarray}
where 
$|x\rangle\equiv\bigotimes_{j=1}^N|x_j\rangle$,
$|z\rangle\equiv\bigotimes_{j=1}^N|z_j\rangle$,
$|k\rangle\equiv\bigotimes_{j=1}^N|k_j\rangle$,
and
\begin{eqnarray*}
{\mathcal H}_Z
\equiv\sum_{i<j}p_{i,j}\frac{I^{\otimes N}+s_{i,j}Z_i\otimes Z_j}{2}.
\end{eqnarray*}

On the other hand,
let us consider Protocol~\ref{protocol:MA}.
For any $x\in A_{yes}$, the verifier's acceptance probability 
$q_{acc}^{honest}(x)$ 
of Protocol~\ref{protocol:MA} is 
\begin{eqnarray*}
q_{acc}^{honest}(x)&=&p_{succ}\frac{1}{2}\sum_{h\in\{0,1\}}\sum_{m\in\{0,1\}^N}
|\langle m|(H^{\otimes N})^h|E_0\rangle|^2
\sum_{i<j}p_{i,j}\frac{1-(-1)^{m_i+m_j}s_{i,j}}{2}\\
&=&
p_{succ}\frac{1}{2}\sum_{h\in\{0,1\}}\sum_{m\in\{0,1\}^N}
\langle m|(H^{\otimes N})^h|E_0\rangle
\langle E_0|(H^{\otimes N})^h
\sum_{i<j}p_{i,j}\frac{I^{\otimes N}-s_{i,j}Z_i\otimes Z_j}{2}|m\rangle\\
&=&
p_{succ}\frac{1}{2}\sum_{h\in\{0,1\}}
\mbox{Tr}\Big[
|E_0\rangle
\langle E_0|(H^{\otimes N})^h
\sum_{i<j}p_{i,j}\frac{I^{\otimes N}-s_{i,j}Z_i\otimes Z_j}{2}
(H^{\otimes N})^h
\Big]\\
&=&
p_{succ}\mbox{Tr}\Big[
|E_0\rangle\langle E_0|
(I^{\otimes N}-{\mathcal H})
\Big]\\
&\ge&p_{acc}^{honest}(x)-\epsilon,
\end{eqnarray*}
where the last inequality is from 
Eq.~(\ref{yes}).

For any $x\in A_{no}$,
the malicious prover samples $m$ from
any probability distribution $D$.
The verifier's acceptance probability $q_{acc}^{malicious}(x)$ is
\begin{eqnarray*}
q_{acc}^{malicious}(x)
&=&p_{succ}\sum_{m\in\{0,1\}^N}
D(m)
\sum_{i<j}p_{i,j}
\frac{1-(-1)^{m_i+m_j}s_{i,j}}{2}\\
&=&p_{succ}\sum_{m\in\{0,1\}^N}
D(m)
\langle m|\sum_{i<j}p_{i,j}
\frac{I^{\otimes N}-s_{i,j}Z_i\otimes Z_j}{2}|m\rangle\\
&=&p_{succ}\mbox{Tr}\Big[(I^{\otimes N}-{\mathcal H}_Z)
\sum_mD(m)|m\rangle\langle m|\Big]\\
&=&
p_{acc}^{malicious}(x),
\end{eqnarray*}
where $|m\rangle\equiv\bigotimes_{j=1}^N|m_j\rangle$ 
and the last equality is from Eq.~(\ref{no}).
Therefore, if $p_{acc}^{honest}$ and
$p_{acc}^{malicious}$ have $\frac{1}{poly(|x|)}$ gap,
and $\epsilon$ is sufficiently small,
then
$q_{acc}^{honest}$ and
$q_{acc}^{malicious}$ also have $\frac{1}{poly(|x|)}$ gap,
which means $A$ is in ${\rm MA}_{\rm BQP}$.
Hence we have shown that ${\rm BQP}\subseteq{\rm MA}_{\rm BQP}$.
\fbox

\begin{figure}[h]
\rule[1ex]{\textwidth}{0.5pt}
\begin{itemize}
\item[1.]
If the prover is honest, it uniformly randomly chooses
$h\in\{0,1\}$, generates a low-energy state
$|E_0\rangle$ of the local Hamiltonian ${\mathcal H}$,
and measures each qubit of $|E_0\rangle$ in the computational
(Hadamard) basis if $h=0$ $(h=1)$.
The prover sends $m\equiv(m_1,...,m_N)\in\{0,1\}^N$ to the verifier,
where
$m_i$ is the measurement result on the $i$th qubit.
If the prover is malicious, the prover sends any $m$ to the verifier.
\item[2.]
The verifier rejects with probability $1-p_{succ}$.
With probability $p_{succ}$,
the verifier samples $(i,j)$ with probability $p_{i,j}$,
and accepts if and only if $(-1)^{m_i+m_j}=-s_{i,j}$.
\end{itemize} 
\rule[1ex]{\textwidth}{0.5pt}
\caption{The ${\rm MA}_{\rm BQP}$ protocol.}
\label{protocol:MA}
\end{figure}

\section{More general setup}
\label{sec:result2}
In this section, we study a more general setup and show a similar
no-go result. Let us consider the verification protocol,
Protocol~\ref{protocol:generalQ}.
In the first step, the verifier (or the trusted center) generates
quantum states $\{\rho_i\}_i$.
We assume that this quantum process is a simple one
(for example, $\rho_i$ is an $N$-tensor product of random
BB84 states), because the verifier's (or the trusted center's)
quantum burden should be minimum.
(If the verifier can do complicated quantum computing, 
there is no point in delegating quantum computing to the prover:
the verifier can do the quantum computation by itself.
Furthermore, if a trusted center that can do complicated quantum computing
is available, the verifier has only to use it
instead of interacting with the untrusted prover.)

We show that the first quantum message transmission (step 1) 
of Protocol~\ref{protocol:generalQ} cannot be replaced with
a ccRSP protocol unless ${\rm BQP}\subseteq{\rm IP}_{\rm BQP}[const]$,
where ${\rm IP}_{\rm BQP}[const]$ is 
the IP with a constant round and a honest quantum polynomial-time
prover.
Because
${\rm IP}_{\rm BQP}[const]
\subseteq{\rm IP}[const]
\subseteq{\rm AM}$,
it means
${\rm BQP}\subseteq{\rm AM}$.

Let us consider Protocol~\ref{protocol:generalC}
that is equivalent to Protocol~\ref{protocol:generalQ}
except that the first quantum step of 
Protocol~\ref{protocol:generalQ}
is replaced with a ccRSP protocol.
We consider a general setup where
the ccRSP protocol is an approximate one:
even if the prover is honest, they succeed with probability $p_{succ}$,
and what the prover gets is a state $\rho_i'$ with probability $p_i'$,
where $\rho_i'$ is close to $\rho_i$ and
$\{p_i'\}_i$ is close to $\{p_i\}_i$.
Furthermore, we assume that $p_{succ}$ is known,
$\{p_i'\}_i$ is samplable in classical polynomial-time,
and $\rho_i'$ can be generated in quantum polynomial-time.
These assumptions are reasonable, because the description of the
ccRSP protocol is known to the verifier, and
$\{\rho_i'\}_i$ and $\{p_i'\}_i$ are close
to
$\{\rho_i\}_i$ and $\{p_i\}_i$, respectively.

\begin{theorem}
Assume that Protocol~\ref{protocol:generalC}
can verify any BQP problem.
It means that
for any promise problem $A=(A_{yes},A_{no})$ in BQP,
Protocol~\ref{protocol:generalC} satisfies both of the following
with some $c$ and $s$ such that $c-s\ge\frac{1}{poly(|x|)}$:
\begin{itemize}
\item
If $x\in A_{yes}$, the honest quantum polynomial-time prover's
behavior makes the verifier
accept with probability at least $c$.
\item
If $x\in A_{no}$,
the verifier's acceptance probability is at most $s$ for any
(even unbounded)
prover's deviation.
\end{itemize}
Then, ${\rm BQP}\subseteq{\rm IP}_{\rm BQP}[const]$.
\end{theorem}

Remark. Again, the theorem requires only the correctness for
the ccRSP. Neither the blindness nor the verifiability is required.

\begin{figure}[h]
\rule[1ex]{\textwidth}{0.5pt}
\begin{itemize}
\item[1.]
The verifier generates a state $\rho_i$ with probability $p_i$,
and sends it to the prover.
Or, the trusted center generates a state $\rho_i$ with probability $p_i$,
sends it to the prover, and sends its classical description
$[\rho_i]$ to the verifier.
\item[2.]
The prover and the verifier exchange a constant round of classical
messages.
The honest prover is quantum polynomial-time, but the malicious
prover is unbounded.
The verifier is classical probabilistic polynomial-time.
\item[3.]
The verifier finally makes the decision.
\end{itemize} 
\rule[1ex]{\textwidth}{0.5pt}
\caption{The general protocol with quantum channel.}
\label{protocol:generalQ}
\end{figure}

\begin{figure}[h]
\rule[1ex]{\textwidth}{0.5pt}
\begin{itemize}
\item[1.]
The prover and the verifier run a ccRSP protocol.
If the prover is honest, with probability $p_{succ}$,
the prover gets a state $\rho_i'$ with
probability $p_i'$, and the verifier gets the classical description 
$[\rho_i']$ of $\rho_i'$.
With probability $1-p_{succ}$, they fail, and the prover and the verifier
get an error message.
If they fail, the verifier rejects.
\item[2.]
The same as the step 2 of Protocol~\ref{protocol:generalQ}.
\item[3.]
The same as the step 3 of Protocol~\ref{protocol:generalQ}.
\end{itemize} 
\rule[1ex]{\textwidth}{0.5pt}
\caption{The general protocol with ccRSP.}
\label{protocol:generalC}
\end{figure}

{\it Proof.}
Let $A=(A_{yes},A_{no})$ be any BQP promise problem.
For any yes instance $x\in A_{yes}$, 
let $p_{acc}^{honest}(x)$ be
the verifier's acceptance probability 
when the prover is honest in Protocol~\ref{protocol:generalC}.

For any no instance $x\in A_{no}$,
let us consider the following malicious unbounded prover's attack
in Protocol~\ref{protocol:generalC}:
\begin{itemize}
\item[1.]
When the prover and the verifier run the ccRSP protocol,
the prover classically simulates prover's honest quantum behavior.
(See Appendix~\ref{app:learn}.)
If they succeed,
the verifier gets $[\rho_i']$ with probability $p_i'$.
The prover can learn $[\rho_i']$,
because the prover has the classical description of $\rho_i'$.
(See Appendix~\ref{app:learn}.)
\item[2.]
When the prover and the verifier exchange classical messages,
the prover does any malicious behavior.
\end{itemize}

Let us consider Protocol~\ref{protocol:AM}.
For any yes instance $x\in A_{yes}$,
let $q_{acc}^{honest}(x)$ be
the verifier's acceptance probability 
with the honest prover in Protocol~\ref{protocol:AM}.
Obviously,
\begin{eqnarray}
p_{acc}^{honest}(x)=q_{acc}^{honest}(x).
\end{eqnarray}

For any no instance $x\in A_{no}$,
let $q_{acc}^{malicious}(x)$ be
the verifier's acceptance probability 
in Protocol~\ref{protocol:AM} with the malicious prover.
It is also easy to see that
\begin{eqnarray}
p_{acc}^{malicious}(x)=q_{acc}^{malicious}(x).
\end{eqnarray}
Therefore, if Protocol~\ref{protocol:generalC}
can verify the promise problem $A$,
Protocol~\ref{protocol:AM} can also verify it,
which means that $A$ is in ${\rm IP}_{\rm BQP}[const]$.
\fbox

\begin{figure}[h]
\rule[1ex]{\textwidth}{0.5pt}
\begin{itemize}
\item[1.]
With probability $p_{succ}$,
the verifier chooses $i$ with probability $p_i'$
and sends $i$ to the prover.
If the prover is honest, it generates $\rho_i'$.
With probability $1-p_{succ}$, the verifier rejects.
\item[2.]
The same as the step 2 of
Protocol~\ref{protocol:generalC}.
\item[3.]
The same as the step 3 of
Protocol~\ref{protocol:generalC}.
\end{itemize} 
\rule[1ex]{\textwidth}{0.5pt}
\caption{The ${\rm IP}_{\rm BQP}[const]$ protocol.}
\label{protocol:AM}
\end{figure}

\section{Computational soundness}
\label{sec:computational}
We have seen that the replacement of the
trusted center in the protocol of Ref.~\cite{TC}
with the ccRSP
does not realize the information-theoretically-sound
classical verification of quantum computing.
What happens if we consider the computational soundness?
In this section, we show that if a ccRSP protocol satisfies a certain
condition,
the protocol of Ref.~\cite{TC} with the ccRSP
is the classical verification of quantum computing
(with the computational soundness).

\begin{theorem}
\label{theorem:computational}
Assume that a ccRSP protocol satisfies the following:
For any quantum polynomial-time malicious
prover's deviation, the verifier gets
$(h,m)\in\{0,1\}^{N+1}$
with probability
\begin{eqnarray*}
P(h,m)&\equiv&\frac{1}{2}\mbox{Tr}
\Big[
(I^{\otimes M}_{B_1}\otimes|\phi_{h,m}\rangle\langle\phi_{h,m}|_{B_2})
\rho_{B_1,B_2}
(I^{\otimes M}_{B_1}\otimes|\phi_{h,m}\rangle\langle\phi_{h,m}|_{B_2})
\Big],
\end{eqnarray*}
and
the prover gets a state 
\begin{eqnarray*}
\sigma_{h,m}\equiv
\frac{1}{2P(h,m)}\mbox{Tr}_{B_2}
\Big[
(I^{\otimes M}_{B_1}\otimes|\phi_{h,m}\rangle\langle\phi_{h,m}|_{B_2})
\rho_{B_1,B_2}
(I^{\otimes M}_{B_1}\otimes|\phi_{h,m}\rangle\langle\phi_{h,m}|_{B_2})
\Big]
\end{eqnarray*}
(up to a CPTP map on it),
where 
$B_1$ is a subsystem of $M$ qubits,
$B_2$ is a subsystem of $N$ qubits,
$|\phi_{h,m}\rangle\equiv\bigotimes_{j=1}^NH^h|m_j\rangle$,
$\rho_{B_1,B_2}$ is any $(M+N)$-qubit state (that could be
chosen by the prover), and
$\mbox{Tr}_{B_2}$ is the partial trace over the subsystem
$B_2$. 
Then, if we replace the trusted center of the protocol of Ref.~\cite{TC}
with the ccRSP protocol, it is the classical verification of
quantum computing (with the computational soundness).
\end{theorem}

Before showing the theorem, we have three remarks.
First, note that when
\begin{eqnarray*}
\rho_{B_1,B_2}=\Big(
\frac{|00\rangle+|11\rangle}{\sqrt{2}}
\frac{\langle00|+\langle11|}{\sqrt{2}}
\Big)^{\otimes N},
\end{eqnarray*}
$P(h,m)=\frac{1}{2^{N+1}}$ for any $(h,m)$ and
$\sigma_{h,m}=\bigotimes_{j=1}^N H^h|m_j\rangle\langle m_j|H^h$,
which corresponds to
the honest prover case.

Second, the above condition is not satisfied against the unbounded malicious prover,
because, as is shown in Appendix~\ref{app:learn}, the unbounded malicious
prover can get the classical description of $\sigma_{h,m}$
and therefore what the prover gets is not $\sigma_{h,m}$
but, for example, $\sigma_{h,m}\otimes|h,m\rangle\langle h,m|$.

Third, it was believed that the verifiability is necessary for a
ccRSP protocol when it is used as a subroutine of the verification
of quantum computing: even if malicious Bob deviates during
the ccRSP protocol,
it should be guaranteed that
the correct state is generated in Bob's place
(up to his operation on it).
Theorem~\ref{theorem:computational} 
suggests that it is not necessarily the case:
as long as it is guaranteed that Bob does the correct measurement
(i.e., the projection $|\phi_{h,m}\rangle\langle \phi_{h,m}|$ )
on any state, the soundness of the verification protocol holds.
It is easy to see that
the verifiability is a special case of our condition: 
In our condition, $\rho_{B_1,B_2}$ is any, but the verifiability
requires that $\rho_{B_1,B_2}$ is the $N$-tensor product
of the Bell pair.
Our condition is therefore weaker than the verifiability.

{\it Proof.}
Let $A=(A_{yes},A_{no})$ be any promise problem in BQP.
The completeness is obvious.
For any yes instance $x\in A_{yes}$, it is clear that
the verifier's acceptance probability with the honest prover is
$
p_{acc}=1-\mbox{Tr}(|E_0\rangle\langle E_0|{\mathcal H})\ge1-\alpha.
$
(See Appendix~\ref{app:TC}.)

Let us next consider the soundness. 
The verifier's acceptance probability $p_{acc}$ with the malicious
prover is
\begin{eqnarray*}
p_{acc}&=&
\sum_{h,m}P(h,m)
\sum_{x,z}\mbox{Tr}(\Pi_{x,z}\sigma_{h,m})
\sum_{i<j}p_{i,j}\frac{1-s_{i,j}(-1)^{m_i'+m_j'}}{2}\\
&=&
\sum_{h,m}P(h,m)
\sum_{x,z}\frac{1}{2P(h,m)}
\mbox{Tr}\Big[\Big(\Pi_{x,z}\otimes 
|\phi_{h,m}\rangle\langle\phi_{h,m}|\Big)
\rho_{B_1,B_2}\Big]
\sum_{i<j}p_{i,j}\frac{1-s_{i,j}(-1)^{m_i'+m_j'}}{2}\\
&=&
\frac{1}{2}\sum_{h,m}
\sum_{x,z}
\mbox{Tr}\Big[\rho_{B_1,B_2}\\
&&\Big\{\Pi_{x,z}\otimes 
(H^{\otimes N})^h
|m\rangle\langle m|
X^{hz+(1-h)x}
\sum_{i<j}p_{i,j}\frac{I^{\otimes N}-s_{i,j}Z_i\otimes Z_j}{2}
X^{hz+(1-h)x}(H^{\otimes N})^h
\Big\}
\Big]\\
&=&
\frac{1}{2}
\sum_h
\sum_{x,z}
\mbox{Tr}\Big[\rho_{B_1,B_2}\\
&&\Big\{\Pi_{x,z}\otimes 
(H^{\otimes N})^h
X^{hz+(1-h)x}
\sum_{i<j}p_{i,j}\frac{I^{\otimes N}-s_{i,j}Z_i\otimes Z_j}{2}
X^{hz+(1-h)x}(H^{\otimes N})^h
\Big\}
\Big]\\
&=&
\sum_{x,z}
\mbox{Tr}\Big[\rho_{B_1,B_2}
\Big\{\Pi_{x,z}\otimes 
Z^zX^x(I^{\otimes N}-{\mathcal H})X^xZ^z
\Big\}
\Big]\\
&=&1-\mbox{Tr}({\mathcal H}\eta)\\
&\le&1-\beta,
\end{eqnarray*}
where
$X^x\equiv\bigotimes_{j=1}^N X^{x_j}$,
$Z^z\equiv\bigotimes_{j=1}^N Z^{z_j}$,
$X^{hz+(1-h)x}\equiv\bigotimes_{j=1}^N X^{hz_j+(1-h)x_j}$,
$|m\rangle\equiv\bigotimes_{j=1}^N|m_j\rangle$,
and
\begin{eqnarray*}
\eta\equiv
\mbox{Tr}_{B_1}
\Big[\sum_{x,z}
(\sqrt{\Pi_{x,z}}\otimes X^xZ^z)
\rho_{B_1,B_2}
(\sqrt{\Pi_{x,z}}\otimes Z^zX^x)
\Big]
\end{eqnarray*}
is an $N$-qubit state.
\fbox

\section{Off-line-quantum communication from prover to verifier}
\label{sec:offline}

The trusted center model~\cite{TC} (Protocol~\ref{protocol:TC}) does
not need any quantum communication between the prover and the verifier.
The FK protocol requires quantum communication from the verifier to
the prover.
The posthoc protocol~\cite{posthoc} requires
quantum communication from the prover to the verifier.
A difference between the FK protocol and the posthoc protocol is that
the FK protocol is off-line-quantum but the posthoc protocol
is on-line-quantum. It means that in the FK protocol, the first quantum
message from the verifier to the prover is independent of the instance
that the verifier wants to verify, but in the posthoc protocol,
the quantum message from the prover to the verifier depends on the
instance.

Is it possible to construct a verification protocol with off-line-quantum
communication from the prover to the verifier?
Theorem~\ref{theorem:computational} answers to the question.
The condition of Theorem~\ref{theorem:computational} is satisfied
when the prover generates a quantum state $\rho_{B_1,B_2}$,
and sends $B_2$ register to the verifier.
Let us consider Protocol~\ref{protocol:offline}. 
From Theorem~\ref{theorem:computational}, it is easy to see that
the protocol is a verification protocol
with off-line-quantum communication from the prover to the verifier.

\begin{figure}[h]
\rule[1ex]{\textwidth}{0.5pt}
\begin{itemize}
\item[0.]
The same as the step 0 of Protocol~\ref{protocol:TC}.
\item[1.]
The prover generates a state $\rho_{B_1,B_2}$ and sends
the register $B_2$ to the verifier.
If the prover is honest, $\rho_{B_1,B_2}$ is the $N$-tensor-product
of Bell pairs.
\item[2.]
The verifier uniformly randomly chooses $h\in\{0,1\}$.
If $h=0$ ($h=1$) the verifier measures each qubit sent from the prover
in the computational (Hadamard) basis.
Let $m_j\in\{0,1\}$ 
be the measurement result on the $j$th qubit $(j=1,2,...,N)$.
\item[3.]
The same as the steps 2 and 3 of Protocol~\ref{protocol:TC}.
\end{itemize} 
\rule[1ex]{\textwidth}{0.5pt}
\caption{The off-line-quantum prover-to-verifier protocol.}
\label{protocol:offline}
\end{figure}

\section{Extractors}
\label{sec:extractor}
In this section, we show that the trusted center verification
protocol of Ref.~\cite{TC} and its variant with the ccRSP studied
in Sec.~\ref{sec:computational} have extractors for
low-energy states.

\begin{theorem}
\label{theorem:pk}
The protocol of Ref.~\cite{TC}
has a quantum polynomial-time extractor that satisfies the following.
When a prover $P^*$ makes the verifier accept an instance $x\in A$ with
probability at least $1-\epsilon$,
the extractor that oracle accesses to $P^*$
outputs a state $\eta$ whose expectation energy 
$\mbox{Tr}(\eta{\mathcal H})$
on the local Hamiltonian ${\mathcal H}$ corresponding to $x$
is less than $\epsilon$.
\end{theorem}

{\it Proof.}
The verifier's acceptance probability $p_{acc}$ against
the prover $P^*$ whose POVM measurement is $\{\Pi_{x,z}\}_{x,z}$
is
\begin{eqnarray*}
p_{acc}&=&\frac{1}{2^{N+1}}\sum_{h,m}
\sum_{x,z}\mbox{Tr}\Big[\Pi_{x,z}(H^{\otimes N})^h|m\rangle\langle m|
(H^{\otimes N})^h\Big]
\sum_{i<j}p_{i,j}\frac{1-s_{i,j}(-1)^{m_i'+m_j'}}{2}\\
&=&\frac{1}{2^{N+1}}\sum_m
\sum_{x,z}
\langle m|
\Pi_{x,z}
Z^zX^x
\sum_{i<j}p_{i,j}\frac{I^{\otimes N}-s_{i,j}Z_i\otimes Z_j}{2}
X^xZ^z
|m\rangle\\
&&+\frac{1}{2^{N+1}}\sum_m
\sum_{x,z}
\langle m|
H^{\otimes N}
\Pi_{x,z}
Z^zX^x
\sum_{i<j}p_{i,j}\frac{I^{\otimes N}-s_{i,j}X_i\otimes X_j}{2}
X^xZ^z
H^{\otimes N}|m\rangle\\
&=&1-\mbox{Tr}[{\mathcal H}\eta],
\end{eqnarray*}
where $|m\rangle\equiv\bigotimes_{j=1}^N|m_j\rangle$,
$X^x\equiv\bigotimes_{j=1}^NX^{x_j}$,
$Z^z\equiv\bigotimes_{j=1}^NZ^{z_j}$,
and
\begin{eqnarray*}
\eta\equiv\frac{1}{2^N}\sum_{x,z}Z^zX^x\Pi_{x,z}X^xZ^z
\end{eqnarray*}
is an $N$-qubit state.

Assume that $p_{acc}\ge1-\epsilon$. Then,
$\mbox{Tr}({\mathcal H}\eta)\le \epsilon$. 
The extractor that outputs $\eta$ can be constructed
in the following way.
The extractor first generates
$\frac{I^{\otimes N}}{2^N}$.
It then does the POVM measurement
$\{\Pi_{x,z}\}_{x,z}$ to obtain the post-measurement state
\begin{eqnarray*}
\sum_{x,z}
\sqrt{\Pi_{x,z}}\frac{I^{\otimes N}}{2^N}
\sqrt{\Pi_{x,z}}\otimes |x,z\rangle\langle x,z|.
\end{eqnarray*}
After the application of the controlled-$XZ$ operation and the 
tracing out of the second register, the extractor obtains
$\eta$.
\fbox

\begin{theorem}
\label{theorem:ccRSP}
Assume that a ccRSP protocol satisfies
the conditions of 
Theorem~\ref{theorem:computational},
and $\rho_{B_1,B_2}$ can be generated in quantum polynomial-time.
Then, 
the protocol of Ref.~\cite{TC}
with the ccRSP
has a quantum polynomial-time extractor that satisfies the following.
When a prover $P^*$ makes the verifier accept an instance $x\in A$ with
probability at least $1-\epsilon$,
the extractor that oracle accesses to $P^*$
outputs a state $\eta$ whose expectation energy 
$\mbox{Tr}(\eta{\mathcal H})$
on the local Hamiltonian ${\mathcal H}$ corresponding to $x$
is less than $\epsilon$.
\end{theorem}

{\it Proof.}
The verifier's acceptance probability is
\begin{eqnarray*}
p_{acc}&=&\sum_{h,m}P(h,m)
\sum_{x,z}\mbox{Tr}(\Pi_{x,z}\sigma_{h,m})
\sum_{i<j}p_{i,j}
\frac{1-s_{i,j}(-1)^{m_i'+m_j'}}{2}\\
&=&
1-\mbox{Tr}({\mathcal H}\eta),
\end{eqnarray*}
where 
$\eta$ is the $N$-qubit state defined by
\begin{eqnarray*}
\eta\equiv
\mbox{Tr}_{B_1}
\Big[
\sum_{x,z}
(\sqrt{\Pi_{x,z}}\otimes X^xZ^z)
\rho_{B_1,B_2}
(\sqrt{\Pi_{x,z}}\otimes Z^zX^x)
\Big].
\end{eqnarray*}

The extractor that outputs $\eta$ can be constructed
in the following way.
The extractor does the POVM measurement on the $B1$ register of
$\rho_{B_1,B_2}$ to generate
\begin{eqnarray*}
\sum_{x,z}
(\sqrt{\Pi_{x,z}}\otimes I^{\otimes N})
\rho_{B_1,B_2}
(\sqrt{\Pi_{x,z}}\otimes I^{\otimes N})
\otimes
|x,z\rangle\langle x,z|.
\end{eqnarray*}
The extractor then applies the controlled-$XZ$ and tracing
out the $B_1$ register and the third register to obtain $\eta$.
\fbox

\appendix
\section{Proof of completeness and soundness}
\label{app:TC}
In this Appendix, we show the completeness and the soundness
of Protocol~\ref{protocol:TC}.
First, we show the completeness.
Let us define the Bell basis by
$
|\phi_{\alpha,\beta}\rangle\equiv
(Z^\beta\otimes X^\alpha)
\frac{|0\rangle\otimes|0\rangle+|1\rangle\otimes|1\rangle}{\sqrt{2}},
$
where $\alpha,\beta\in\{0,1\}$.
We also define
$|m\rangle\equiv\bigotimes_{j=1}^N|m_j\rangle$
and
$X^{hz+(1-h)x}\equiv\bigotimes_{j=1}^N X^{hz_j+(1-h)x_j}$,
where $h\in\{0,1\}$ and $x,z\in\{0,1\}^N$.
The verifier's acceptance probability with the honest prover is
\begin{eqnarray*}
p_{acc}
&=&\frac{1}{2}\sum_{h\in\{0,1\}}
\frac{1}{2^N}\sum_{m\in\{0,1\}^N}
\sum_{x\in\{0,1\}^N}
\sum_{z\in\{0,1\}^N}\\
&&\times
\Big(\bigotimes_{j=1}^N\langle \phi_{x_j,z_j}|\Big)
\Big[|E_0\rangle\langle E_0|\otimes
(H^{\otimes N})^h|m\rangle\langle m|(H^{\otimes N})^h
\Big]\Big(\bigotimes_{j=1}^N|\phi_{x_j,z_j}\rangle\Big)\\
&&\times\sum_{i<j}p_{i,j}\frac{1-s_{i,j}(-1)^{m_i'+m_j'}}{2}\\
&=&\frac{1}{2}\sum_h
\frac{1}{2^N}\sum_m
\sum_{x,z}
\frac{1}{2^N}
\langle m|(H^{\otimes N})^h
Z^zX^x
|E_0\rangle\langle E_0|
X^xZ^z(H^{\otimes N})^h|m\rangle\\
&&\times\sum_{i<j}p_{i,j}\frac{1-s_{i,j}(-1)^{m_i'+m_j'}}{2}\\
&=&\frac{1}{2}\sum_h
\frac{1}{2^N}\sum_m
\sum_{x,z}
\frac{1}{2^N}
\langle m|(H^{\otimes N})^h
Z^zX^x
|E_0\rangle\langle E_0|
X^xZ^z(H^{\otimes N})^h\\
&&\times
X^{hz+(1-h)x}
\sum_{i<j}p_{i,j}\frac{I^{\otimes N}-s_{i,j}Z_i\otimes Z_j}{2}
X^{hz+(1-h)x}
|m\rangle\\
&=&\frac{1}{2}\sum_h
\frac{1}{2^N}
\sum_{x,z}
\frac{1}{2^N}
\mbox{Tr}\Big[
(H^{\otimes N})^h
Z^zX^x
|E_0\rangle\langle E_0|
X^xZ^z(H^{\otimes N})^h\\
&&\times
X^{hz+(1-h)x}
\sum_{i<j}p_{i,j}\frac{I^{\otimes N}-s_{i,j}Z_i\otimes Z_j}{2}
X^{hz+(1-h)x}
\Big]\\
&=&
\frac{1}{2^N}
\sum_{x,z}
\frac{1}{2^N}
\mbox{Tr}\Big[
|E_0\rangle\langle E_0|
\frac{1}{2}\sum_h
X^xZ^z(H^{\otimes N})^h\\
&&\times
X^{hz+(1-h)x}
\sum_{i<j}p_{i,j}\frac{I^{\otimes N}-s_{i,j}Z_i\otimes Z_j}{2}
X^{hz+(1-h)x}
(H^{\otimes N})^h
Z^zX^x
\Big]\\
&=&\mbox{Tr}\Big[
|E_0\rangle\langle E_0|(I^{\otimes N}-{\mathcal H})
\Big]\\
&\ge&1-\alpha.
\end{eqnarray*}
Here, in the second equality, we have used the following result:
for any $\alpha,\beta,h,m\in\{0,1\}$ and any single-qubit state
$\rho$, 
\begin{eqnarray*}
\langle\phi_{\alpha,\beta}|
(\rho\otimes H^h|m\rangle\langle m|H^h)
|\phi_{\alpha,\beta}\rangle
=\frac{1}{2}
\langle m|H^h
Z^\beta X^\alpha
\rho X^\alpha Z^\beta
H^h|m\rangle.
\end{eqnarray*}

Next we show the soundness.
Let $\{\Pi_{x,z}\}_{x,z}$ be the POVM that the malicious prover applies.
The verifier's acceptance probability is
\begin{eqnarray*}
p_{acc}
&=&\frac{1}{2}\sum_h\frac{1}{2^N}\sum_m
\sum_{x,z}\langle m|(H^{\otimes N})^h
\Pi_{x,z}
(H^{\otimes N})^h|m\rangle
\sum_{i<j}p_{i,j}
\frac{1-s_{i,j}(-1)^{m_i'+m_j'}}{2}\\
&=&\frac{1}{2}\sum_h\frac{1}{2^N}\sum_m
\sum_{x,z}\langle m|(H^{\otimes N})^h
\Pi_{x,z}
(H^{\otimes N})^h
X^{hz+(1-h)x}\\
&&\times\sum_{i<j}p_{i,j}
\frac{I^{\otimes N}-s_{i,j}Z_i\otimes Z_j}{2}
X^{hz+(1-h)x}
(H^{\otimes N})^h
(H^{\otimes N})^h|m\rangle\\
&=&\frac{1}{2}\sum_h\frac{1}{2^N}
\sum_{x,z}
\mbox{Tr}\Big[(H^{\otimes N})^h
\Pi_{x,z}
(H^{\otimes N})^h
X^{hz+(1-h)x}\\
&&\times\sum_{i<j}p_{i,j}
\frac{I^{\otimes N}-s_{i,j}Z_i\otimes Z_j}{2}
X^{hz+(1-h)x}
(H^{\otimes N})^h
(H^{\otimes N})^h\Big]\\
&=&\mbox{Tr}\Big[
(I^{\otimes N}-{\mathcal H})\sigma\Big]\\
&\le&1-\beta,
\end{eqnarray*}
where
$
\sigma\equiv
\frac{1}{2^N}\sum_{x,z}X^xZ^z\Pi_{x,z}Z^zX^x,
$
and the last inequality is from the fact that
$\sigma$ is a state
because
$\mbox{Tr}(\sigma)=1$
and $\sigma\ge0$.
\fbox

\section{Unbounded prover can learn $(h,m)$}
\label{app:learn}
In this Appendix, we show that the unbounded malicious prover can
learn $(h,m)$.
Without loss of generality, a ccRSP protocol when
the prover is honest is described as follows:
\begin{itemize}
\item[1.]
The verifier sends a classical message $a_1$ to the prover.
\item[2.]
The prover generates a state $\rho_1(a_1)$.
\item[3.]
The prover measures some qubits of $\rho_1(a_1)$ in the computational
basis to obtain a result $b_1$.
The prover sends $b_1$ to the verifier.
Let $\rho_1'(a_1,b_1)$ be the post-measurement state.
\item[4.]
The verifier sends a classical message $a_2$ to the prover.
\item[5.]
The prover applies a unitary on $\rho_1'(a_1,b_1)$ 
to generate a state $\rho_2(a_1,b_1,a_2)$.
The prover measures some qubits of
$\rho_2(a_1,b_1,a_2)$ in the computational basis to obtain a result $b_2$.
The prover sends $b_2$ to the verifier.
Let $\rho_2'(a_1,b_1,a_2,b_2)$ be the post-measurement state.
\item[6.]
The verifier sends a classical message $a_3$ to the prover.\\
...
\item[k.]
The verifier outputs $(h,m)\in\{0,1\}^{N+1}$.
The prover has a state $\sigma_{h,m}\otimes\rho_{junk}$.
\end{itemize}
The unbounded prover can simulate the above process classically
as follows:
\begin{itemize}
\item[1.]
The verifier sends a classical message $a_1$ to the prover.
\item[2.]
The prover classically computes the classical description of  
$\rho_1(a_1)$.
\item[3.]
The prover classically samples $b_1$ with probability
$\mbox{Tr}[(|b_1\rangle\langle b_1|\otimes I)\rho_1(a_1)]$.
The prover sends $b_1$ to the verifier.
Let $\rho_1'(a_1,b_1)$ be the post-measurement state.
The prover classically computes the classical description of
$\rho_1'(a_1,b_1)$.
\item[4.]
The verifier sends a classical message $a_2$ to the prover.
\item[5.]
The prover classically computes the classical description of
$\rho_2(a_1,b_1,a_2)$.
The prover classically samples $b_2$ with probability
$\mbox{Tr}[(|b_2\rangle\langle b_2|\otimes I)\rho_2(a_1,b_1,a_2)]$.
The prover sends $b_2$ to the verifier.
Let $\rho_2'(a_1,b_1,a_2,b_2)$ be the post-measurement state.
The prover classically computes the classical description of
$\rho_2'(a_1,b_1,a_2,b_2)$.
\item[6.]
The verifier sends a classical message $a_3$ to the prover.\\
...
\item[k.]
The verifier outputs $(h,m)\in\{0,1\}^{N+1}$.
The prover has a classical description of 
$\sigma_{h,m}\otimes\rho_{junk}$.
\end{itemize}
The verifier cannot distinguish whether the prover is doing
the honest quantum procedure or simulating it classically.
Because the prover has the classical description of
$\sigma_{h,m}$, the prover can learn
$(h,m)$.

\acknowledgements
TM is supported by MEXT Q-LEAP, JST PRESTO No.JPMJPR176A,
the Grant-in-Aid for Young Scientists (B) No.JP17K12637 of JSPS, 
and the Grant-in-Aid for Scientific Research (B) No.JP19H04066 of JSPS.
YT is supported by MEXT Q-LEAP.

\end{document}